\documentstyle[12pt]{article}
\def\tr{\mbox{tr}\,}
\def\e#1{(\ref{#1})}
\newcommand{\be}{\begin{equation}}
\newcommand{\ee}{\end{equation}}
\newcommand{\bea}{\begin{eqnarray}}
\newcommand{\eea}{\end{eqnarray}}

\newcommand{\bbblb}[1]{\setbox\@tempboxa\hbox{$#1[$}%
             \@tempdimb\wd\@tempboxa
             \copy\@tempboxa \kern -.85\@tempdimb
             \copy\@tempboxa \kern -.65\@tempdimb\box\@tempboxa}
\newcommand{\bbbrb}[1]{\setbox\@tempboxa\hbox{$#1]$}%
             \@tempdimb\wd\@tempboxa
             \copy\@tempboxa \kern -.65\@tempdimb
             \copy\@tempboxa \kern -.85\@tempdimb\box\@tempboxa}

\def\e#1{(\ref{#1})}  

\def\bbbone{{\mathchoice {\rm 1\mskip-4mu l} {\rm 1\mskip-4mu l}
{\rm 1\mskip-4.5mu l} {\rm 1\mskip-5mu l}}}
\def\bbbc{{\mathchoice {\setbox0=\hbox{$\displaystyle\rm C$}\hbox{\hbox
to0pt{\kern0.4\wd0\vrule height0.9\ht0\hss}\box0}}
{\setbox0=\hbox{$\textstyle\rm C$}\hbox{\hbox
to0pt{\kern0.4\wd0\vrule height0.9\ht0\hss}\box0}}
{\setbox0=\hbox{$\scriptstyle\rm C$}\hbox{\hbox
to0pt{\kern0.4\wd0\vrule height0.9\ht0\hss}\box0}}
{\setbox0=\hbox{$\scriptscriptstyle\rm C$}\hbox{\hbox
to0pt{\kern0.4\wd0\vrule height0.9\ht0\hss}\box0}}}}
\def\bbbq{{\mathchoice {\setbox0=\hbox{$\displaystyle\rm Q$}\hbox{\raise
0.15\ht0\hbox to0pt{\kern0.4\wd0\vrule height0.8\ht0\hss}\box0}}
{\setbox0=\hbox{$\textstyle\rm Q$}\hbox{\raise
0.15\ht0\hbox to0pt{\kern0.4\wd0\vrule height0.8\ht0\hss}\box0}}
{\setbox0=\hbox{$\scriptstyle\rm Q$}\hbox{\raise
0.15\ht0\hbox to0pt{\kern0.4\wd0\vrule height0.7\ht0\hss}\box0}}
{\setbox0=\hbox{$\scriptscriptstyle\rm Q$}\hbox{\raise
0.15\ht0\hbox to0pt{\kern0.4\wd0\vrule height0.7\ht0\hss}\box0}}}}
\def\bbbt{{\mathchoice {\setbox0=\hbox{$\displaystyle\rm
T$}\hbox{\hbox to0pt{\kern0.3\wd0\vrule height0.9\ht0\hss}\box0}}
{\setbox0=\hbox{$\textstyle\rm T$}\hbox{\hbox
to0pt{\kern0.3\wd0\vrule height0.9\ht0\hss}\box0}}
{\setbox0=\hbox{$\scriptstyle\rm T$}\hbox{\hbox
to0pt{\kern0.3\wd0\vrule height0.9\ht0\hss}\box0}}
{\setbox0=\hbox{$\scriptscriptstyle\rm T$}\hbox{\hbox
to0pt{\kern0.3\wd0\vrule height0.9\ht0\hss}\box0}}}}
\def\bbbs{{\mathchoice
{\setbox0=\hbox{$\displaystyle     \rm S$}\hbox{\raise0.5\ht0\hbox
to0pt{\kern0.35\wd0\vrule height0.45\ht0\hss}\hbox
to0pt{\kern0.55\wd0\vrule height0.5\ht0\hss}\box0}}
{\setbox0=\hbox{$\textstyle        \rm S$}\hbox{\raise0.5\ht0\hbox
to0pt{\kern0.35\wd0\vrule height0.45\ht0\hss}\hbox
to0pt{\kern0.55\wd0\vrule height0.5\ht0\hss}\box0}}
{\setbox0=\hbox{$\scriptstyle      \rm S$}\hbox{\raise0.5\ht0\hbox
to0pt{\kern0.35\wd0\vrule height0.45\ht0\hss}\raise0.05\ht0\hbox
to0pt{\kern0.5\wd0\vrule height0.45\ht0\hss}\box0}}
{\setbox0=\hbox{$\scriptscriptstyle\rm S$}\hbox{\raise0.5\ht0\hbox
to0pt{\kern0.4\wd0\vrule height0.45\ht0\hss}\raise0.05\ht0\hbox
to0pt{\kern0.55\wd0\vrule height0.45\ht0\hss}\box0}}}}
\def\bbbz{{\mathchoice {\hbox{$\textstyle Z\kern-0.5em Z$}}
{\hbox{$\textstyle Z\kern-0.5em Z$}}
{\hbox{$\scriptstyle Z\kern-0.35em Z$}}
{\hbox{$\scriptscriptstyle Z\kern-0.2em Z$}}}}
\def\diameter{{\ifmmode\oslash\else$\oslash$\fi}}
\def\hbar{{\mathchar'26\mkern-9muh}}
\def\qed{\ifmmode\sq\else{\unskip\nobreak\hfil
\penalty50\hskip1em\null\nobreak\hfil\sq
\parfillskip=0pt\finalhyphendemerits=0\endgraf}\fi}

\begin{document}

\footskip=0.6cm
\voffset=.5cm
\textheight=18.5cm
\textwidth=12.5cm
\thispagestyle{empty}
\begin{center}
\null

{\large {\sc Bulgarian Academy of Sciences}\\
\bf  Institute for Nuclear Research and Nuclear Energy\\}
{\normalsize boul. Tzarigradsko shosse 72, 1784 Sofia, Bulgaria\\}
\vskip 3pt
{\scriptsize \bf Tel.: 003592--74311, \hfil Fax.: 003592--755019, \hfil
 Telex: 24368 ISSPH BG \\}
\vskip 2pt
\hrule
\end{center}

\thispagestyle{empty}
\begin{flushright}
hep-th/9502001 \\
{\bf Preprint TH--94/8}
\end{flushright}
\begin{center}
 {\Large \bf Inverse scattering transform analysis of  \\
Stokes-anti-Stokes stimulated Raman scattering }
\end{center}
\begin{center}
V.\ S.\ Gerdjikov
\end{center}
\begin{center}
 {\it Institute of Nuclear Research and Nuclear Energy \\
Blvd.\ Tsarigradsko shosse 72, Sofia 1784, Bulgaria}
\end{center}
\begin{center}
N.\ A.\ Kostov
\end{center}
\begin{center}
{\it Institute of Electronics, Bulgarian Academy of Sciences, \\
Blvd.\ Tsarigradsko shosse 72, Sofia 1784, Bulgaria}
\end{center}

\begin{abstract}
Zakharov-Shabat--Ablowitz-Kaup-Newel-Segur (ZS--AKNS) representation for
Stokes-anti-Stokes stimulated Raman scattering (SRS) is proposed. Periodical
waves, solitons and self-similarity solutions are derived. Transient and
bright threshold solitons are discussed.
\end{abstract}

\vfill
\begin{center} {\bf December 29, 1994 }

{\bf Sofia}
\end{center}
\vfill

\newpage

\section{ Introduction}

The equations that describe the propagation in a Raman active medium when
Stokes $E_s$, anti-Stokes $E_a$ and pump $E_p$ exists and when there is no
frequency mismatch can be written \cite{1,2,3,4,5,6} in the form:
\begin{equation}\label{eq:k1}
\begin{array}{lll} \displaystyle
\frac{\partial E_p}{\partial \zeta}= \beta_a Q^{*} E_a-Q E_s,  &\qquad &
\displaystyle
\frac{\partial E_s}{\partial \zeta} = Q^{*} E_p,  \\[10pt]
\displaystyle \frac{\partial E_a}{\partial \zeta}  = -\beta_a Q E_p, &
\qquad & \displaystyle \frac{\partial Q}{\partial \tau} + \tilde{g} Q=
E_s^{*} E_p+\beta_a E_p^{*} E_a, \end{array}
\end{equation}
where Q is the normalized effective polarization of the medium,
$\zeta=z/L$ and $\tau=t-z/v$ are dimensionless space and retarded time
coordinates, respectively, $v$ is the wave group velocity.  $T_2$ is the
natural damping time of the material excitation and $\tilde{g}=1/T_2$.
By $\beta_a$
we denote the coupling coefficient which determines the number of
anti-Stokes photons relative to number of Stokes photons and its magnitude
depends on the matrix element that describes the dipole transition
\cite{1,but}. In this paper we consider $\beta_a = 1$ \cite{1}, but some
results are valid for $\beta_a\neq 1$. For $\beta_a=0$ i.e.  when the
anti-Stokes wave $E_a$ is neglected, these equations are the so called
transient stimulated Raman scattering (SRS) equations, which possess
ZS--AKNS~\cite{AKNS,22} representation when $\tilde{g}=0$ \cite{7}. The
SRS soliton solutions theoretically discovered by Chu and Scott \cite{7}
have been experimentally observed first in \cite{8}. The SRS solitons
(regarded as transient solitons \cite{9,10,11} with a $\pi$ phase jump at
the Stokes frequency) have been extensively studied \cite{12}. In later
experiments by Duncan {\it et al} \cite{13} a careful comparison between
theory and experiment showed good agreement. Shortly after this work
\cite{13}, Hilfer and Menyuk \cite{14} carried out simulations which
indicate that in the highly depleted regime the solutions of transient SRS
equations always tend  toward a self-similar solution.  This result has
been recovered by Menyuk {\it et al} \cite{15,16} applying the inverse
scattering transform method (ISM) to the transient SRS equations, see
\cite{9}.  Experiment to observe this solution has been proposed
in~\cite{15}.  The Kaup's theory \cite{10} also indicates that the
dissipation, which appears for finite $T_2$ plays a crucial role in
soliton formation. The similarity solutions and other group invariant
solutions of the SRS equations in the presence of dissipation are studied
in \cite{17}.  Claude and Leon \cite{leon} reformulated the transient SRS
equations  as an equivalent $\bar{\partial} $--problem and thus were able
to treat the inhomogeneous broadening with initial conditions of Dr\"{u}hl
{\it et al} \cite{8} experiments.  As a consequence they showed the that
{\it Raman spike} observed in the experiment is not a soliton.

   For general system (\ref{eq:k1}) with anti-Stokes wave, phase
mismatch, $\beta_a\neq 1$ and dissipation, transient $\pi$ solitons have
been investigated by M.~Scalora {\it et al} \cite{6} using numerical
methods. They predict the formation of solitonlike pulses at the
anti-Stokes frequency.  Another type threshold bright $2\pi$
solitons, which have Lorentzian form have been theoretically obtained in
\cite{18}.

   In sect. II we introduce new variables $S_{3}, S_{\pm}$ (\ref{eq:k1q})
which are quadratic in terms of $E_p, E_s, E_a$. Then the system \e{eq:k2}
for $S_{3}$, $S_{\pm}$ derived from (\ref{eq:k1}) with $\beta_a = 1$,
$\tilde{g}=0$ allows ZS-AKNS representation similar to the one
used by Chu and Scott \cite{7} for another physical quantities: difference
of normalized Stokes--anti-Stokes local intensities and for normalized
(complex) local Rabi frequency. We also introduce ``nonlinear time'' and
renormalized dimensionless variables different from yhe ones used
in~\cite{10}. Then we solve the inverse scattering problem (ISP) for the
system \e{eq:L1.6} with the ``nonlinear time'' $\tau' $ restricted to the
finite interval $0 \leq \tau' \leq 1 $; i.e. we derive the corresponding
Gel'fand--Levitan--Marchenko (GLM) equation.

In sect. III  we obtain new periodic, soliton and self--similarity
solutions of Stokes--anti-Stokes SRS equations without dissipation are
obtained for both formulations \e{eq:k1} and \e{eq:k2}. In addition
the transient solitons and the bright solitons of Kaplan {\it et al }
\cite{18} are discussed.

In the last Sect.~IV  we propose an  extension of the Stokes--anti-Stokes
SRS equations for $N$ Stokes and $N$ anti-Stokes waves and conjecture that
it is also  integrable by means of the IST method.

\section{ ZS--AKNS Representation and GLM Equation}

\subsection{ZS--AKNS Representation }
    Let us introduce the following variables
\begin{eqnarray}\label{eq:k1q}
  S_3 = \frac{1}{2} ( |E_s|^2 - |E_a|^2 ), \qquad
S_{+} = \frac{i}{2} ( E_s^{*} E_p + E_p^{*} E_a ), \qquad
S_{+} = S_{-}^{*} .
\end{eqnarray}
In terms of the new quadratic variables (\ref{eq:k1q}) the initial system
with $\tilde{g}=0, \beta_a = 1$ is rewritten as
\begin{eqnarray} \label{eq:k2}
\frac{\partial S_{3}}{\partial \zeta} = -i Q^{*} S_{+}+i Q S_{-} ,
\quad
\frac{\partial S_{+}}{\partial \zeta} = -i Q S_{3} ,
\quad
\frac{\partial Q}{\partial \tau} = -2i S_{+} .
\end{eqnarray}
Then the eq. (\ref{eq:k2}) can be written down as the compatibility
condition
\begin{eqnarray}\label{eq:ZS}
\partial_{\tau} U - \partial_{\zeta} V + [ U,V ] = 0,
\end{eqnarray}
of the following linear systems:
\begin{eqnarray}\label{eq:L1.1}
L(\lambda)F(\zeta,\tau,\lambda) &\equiv &
{\partial F \over \partial \zeta} -
U(\zeta,\tau,\lambda) F(\zeta,\tau,\lambda) =0, \\
\label{eq:L1.1m}
M(\lambda)F(\zeta,\tau,\lambda) &\equiv &
{\partial F \over \partial \tau} -V(\zeta,\tau,\lambda)
F(\zeta,\tau,\lambda) = F(\zeta,\tau,\lambda) C(\lambda),
\end{eqnarray}
where
\begin{eqnarray}\label{eq:u-v}
&& U(\zeta,\tau,\lambda) = -{i\over \lambda} \sigma_3 + {1 \over
\sqrt{2}} q(\zeta,\tau), \qquad V(\zeta,\tau,\lambda) =
{\lambda \over 2i} S(\zeta,\tau),\\
\label{eq:u-v1}
&&q(\zeta,\tau) = \left( \begin{array}{cc} 0 &    Q
\\ -Q^{*} &  0 \end{array} \right) , \qquad S(\zeta,\tau)  =
\left( \begin{array}{ccc} S_3 & -i\sqrt{2} S_{+} \\ i\sqrt{2}
S_{-} & -S_3 \end{array} \right) .
\end{eqnarray}
and $C(\lambda) $ will be fixed up below.

{}From physical point of view \cite{7,8} the initial value problem
associated to the system (\ref{eq:k1}) is the following
\begin{eqnarray}
&&Q(\zeta,0)=0, \qquad  E_{p}(0,\tau)=E_{p0}(\tau), \\
&& E_{s}(0,\tau)=E_{s0}(\tau),  \qquad E_{a}(0,\tau)=E_{a0}(\tau)
\nonumber
\end{eqnarray}
and the problem is to determine the output quantities
$E_{p}(L,\tau)$, $E_{s}(L,\tau)$, $E_{a}(L,\tau)$, where $L$ is the total
length of the beam path in the Raman cell. Analogically the initial value
problem for system (\ref{eq:k2}) is
\begin{eqnarray}
Q(\zeta,0)=0,\quad S_{3}(0,\tau)=S_{30}(\tau), \quad
S_{+}(0,\tau)=S_{+0}(\tau).
\end{eqnarray}

We follow the main idea of~\cite{9,17}, namely that as a Lax operator one
should consider the operator  $M(\lambda) $ in \e{eq:L1.1m} and solve the
inverse scattering problem for it. Then we will use the second operator
$L(\lambda)$  in \e{eq:L1.1} and determine the $\zeta
$--dependence of the corresponding scattering data. However there will be
substancial diferences in the details.

First of all we will approach the inverse scattering problem for the
$M(\lambda)$ operator directly rather than via its gauge equivalence to
a ZS--AKNS type system. Indeed, this equivalence is realized with
$F(\zeta,\tau,\lambda) $ evaluated at $\lambda = 0 $. However, in our case
the other linear problem has a pole singularity at $\lambda =0 $. Due to
fact makes one can not evaluate the $\zeta $--dependence of the gauge
function, which makes impossible the comparizon with the results
in~\cite{9,17}.

It is well known how to solve the ISP for the system \e{eq:L1.1m}
considered on the whole $\tau $--line $-\infty \leq \tau \leq \infty $ and
with boundary conditions of ferromagnetic type, i.e. $\lim_{\tau \to \pm
\infty} S(\zeta,\tau)  = \sigma_3$, see~\cite{22}. We will make use of
these ideas adopting them to our case. First we have to take into account
that the eigenvalues of our $S(\zeta,\tau) $ differ from $\pm 1 $ and are
generically $\tau $--dependent. In order to calculate them it is enough to
know, that $\tr S(\zeta,\tau) =0 $ and
\begin{eqnarray}\label{eq:L1.3}
- \det S(\zeta,\tau) &=& S_3^2 + 2S_+S_- \\
&=& {1 \over 4 } \left( |E_s|^2 - |E_a|^2 \right)^2 + {1 \over 2}
|E_p^* E_a + E_p E_s^*|^2 = K^4(\tau). \nonumber
\end{eqnarray}
Using the evolution equations \e{eq:k1} we check that
\begin{equation}\label{eq:L1.4}
{d K \over d\zeta} = 0.
\end{equation}
Besides from \e{eq:L1.3} we conclude that $K(\tau) $ is real--valued
function. In order to proceed further we require in addition that
$K^2(\tau) $ is monotonic function of $\tau $. Then we can introduce a new
``nonlinear time'' $\tau' $ by
\begin{equation}\label{eq:L1.5}
d \tau ' = K^2(\tau) d \tau
\end{equation}
and the following dimensionless variables:
\begin{eqnarray}\label{eq:L2.2}
\tau'=\int_{0}^{\tau} K^2(\tau'')d\tau''/T_{\infty} ,\qquad
T_{\infty}=\int_{0}^{\infty} K^2(\tau) d\tau ,\qquad
\zeta'=\zeta T_{\infty}\\
E'_p=\frac{E_p}{K(\tau)}, \qquad  E'_s=\frac{E_s}{K(\tau)}, \qquad
E'_a=\frac{E_a}{K(\tau)}, \qquad Q'=\frac{Q}{T_{\infty}} ,\nonumber
\end{eqnarray}
The primed variables introduced above satisfy the same system \e{eq:k1} of
NLEE provided $\tilde{g}'= \tilde{g}T_\infty /K^2 $; in what follows
below we put $\tilde{g}=0 $. Note, that the transformation $\{ E_p, E_s,
E_a, Q\}\rightarrow \{ E'_p, E'_s, E'_a, Q'; K(\tau)\}$ is one--to--one
and invertible. In order to obtain the evolution of $\{ E_p, E_s, E_a,
Q\}$, one must first determine the evolution of $\{ E'_p, E'_s, E'_a,
Q'\}$ and then use the given function $K(\tau)$ to return to the original
variable set.

The nonlinear time $\tau' $ is introduced in analogy to the one
in~\cite{21}; the difference is that now $K^2(\tau) $ can not be
interpreted as the total energy density:
\begin{eqnarray} \label{eq:E}
{\cal  E}(\tau)= |E_p(\zeta,\tau)|^2 +
|E_s(\zeta,\tau)|^2 + |E_a(\zeta,\tau)|^2  ,
\end{eqnarray}
which is constant at every point $\tau$ as function of $\zeta$. Since all
physical solutions possess finite energy we conclude, that each of the
terms in \e{eq:E} must be integrable functions of $\tau $. In particular,
each of these functions must vanish for $\tau \to \infty $. As a
consequence of this fact we conclude that $K^2(\tau) $ must have the same
properties. Therefore for this class of solutions we have $T_\infty <
\infty $ and in terms of $\tau' $ we get the system:
\begin{equation}\label{eq:L1.6}
M'(\lambda)F(\zeta,\tau',\lambda) \equiv {\partial F \over \partial \tau'}
- \lambda S'(\zeta,\tau') F(\zeta,\tau',\lambda) =0 ,
\qquad S'(\zeta,\tau') = {S(\zeta,\tau)  \over  K^2(\tau) }
\end{equation}
where $S'(\zeta,\tau')  $ satisfies $\tr S' =0 $ and $\det S' =-1 $. As a
result the eigenvalues of $S'(\zeta,\tau') $ become equal to $\pm 1 $,
i.e. we can write down
\begin{equation}\label{eq:L2.1}
S'(\zeta,\tau') = g(\zeta,\tau') \sigma_3 g^{-1}(\zeta,\tau')
\end{equation}

Note that the constancy of ${\cal  E}(\tau)$ corresponds to pointwise
conservation of the photon intensity.  From Eq.  (\ref{eq:k1}) it is easy
to derive also the following important relation:
\begin{eqnarray}\label{eq:9}
\frac{1}{2} \frac{\partial}{\partial \tau}  |Q|^2 + \tilde{g} |Q|^2 +
\frac{\partial}{\partial \zeta }  S_3 = 0  .
\end{eqnarray}
from which for $\tilde{g}=0$ we find that $\int |Q|^2 d\zeta$ is an
integral of motion if $S_{3}(0,\tau)-S_{3}(L,\tau)=0$. This will be
fulfilled if $E_a $ and $E_s $ satisfy quasiperiodic boundary conditions,
i.e., if $E_{a,s} (\tau'=0)  = e^{i\phi_{a,s}} E_{a,s} (\tau'=1) $ with
any $\phi_{a,s} $.

In the next subsection we will use only renormalized quantities and the
``nonlinear time'' $\tau' $ and for the simplicity of the notations will
drop all primes.

\subsection{The GLM equation}

We briefly sketch the derivation of the GLM equation related to the
left end $\tau=0 $ of the interval. Of course we have to introduce also
slight modifications in order to take into account the fact that
$S(\tau=0,\zeta) = S_0(\zeta) \neq \sigma_3 $. The operator $M(\lambda)$
on finite interval generically possesses purely discrete spectrum with an
infinite number of simple discrete eigenvalues. As a consequence, the
kernel of the GLM equation contains only a sum over the discrete spectrum.
Skipping the details we write down the results.

Let the Jost solution of \e{eq:L1.6} normalized to the left end $\tau =0 $
of the interval, be fixed up by:
\begin{eqnarray}\label{eq:fas}
\phi_{0} (\tau, \zeta, \lambda) &=& g_0 e^{\lambda \tau \sigma_3 /2i }
, \qquad \lim_{\tau \to 0} \phi (\tau, \zeta,\lambda) =
\lim_{\tau \to 0}\phi_{0} (\zeta, \lambda) = \bbbone , \\
\label{eq:fas1}
\lim_{\tau \to 0} \phi (\tau, \zeta,\lambda) &=&
\lim_{\tau \to 0}g_{1} (\zeta) e^{\lambda\tau\sigma_3/2i}
T(\zeta,\lambda), \\
\label{eq:fas2}
g_0(\zeta) &=& g(\zeta,\tau=0), \qquad g_1(\zeta) = g(\zeta,\tau=1),
\end{eqnarray}
Then its behaviour at $\tau \simeq 1 $ determines the scattering matrix
$T(\zeta,\lambda) $ according to \e{eq:fas1}. It remains now  to evaluate
the ``evolution'' of $T $ in $\zeta $. In order to do this we have to
calculate first $C(\lambda) $ in \e{eq:L1.1m}
 by taking the limit of \e{eq:fas} for $\tau
\to 0 $ with the result $C(\zeta,\lambda) = (i/\lambda) g_0^{-1} \sigma_3
g_0(\zeta) $. Then we take the limit of \e{eq:fas} for $\tau \to 1 $ which
gives the following result for the evolution of $T(\zeta,\lambda) $:
\begin{equation}\label{eq:T-ev}
{dT \over d\zeta } = {i \over \lambda} \left( T(\zeta,\lambda)
g_0^{-1}\sigma_3 g_0 - g_1^{-1}\sigma_3 g_1 T(\zeta,\lambda) \right)
\end{equation}
As we already noted, in order to apply the ISM for the solution of our
problem we will need first to calculate not only $g_0(\zeta) $,  which is
determined from the initial conditions, but also $g_1(\zeta) $. The
situation is greatly simplified if we impose a quasiperiodic boundary
conditions on the fields $E_{a,s,p}(\zeta,\tau) $ in such a way, that
$S_0(\zeta) = S_1(\zeta) $. Then we get $g_0(\zeta)= g_1(\zeta)$ and the
r.h.side of \e{eq:T-ev} becomes proportional to the commutator
$[T,g^{-1}_0(\zeta) \sigma_3 g_0(\zeta)]$. The importance of this
imposition can be seen from the fact, that it immeadiately provides us
with the hierarchy of conservation laws. The generating function of this
hierarchy is $\tr T(\lambda) $, which is now $\zeta $--independent.

Let us also be given $S_0(\zeta) = S(\zeta,\tau =0)$ and let it be
diagonalizable in the form:
\begin{equation}\label{eq:S0}
S_0(\zeta) = g_0(\zeta) \sigma_3 g_0^{-1}(\zeta) , \qquad g_0(\zeta) =
g_1(\zeta,\tau=0).
\end{equation}

We introduce the transformation operator which relates the Jost
solution $\phi (\tau,\zeta,\lambda) $ to its asymptotic
$\phi_{0} (\tau,\zeta,\lambda)  $ \e{eq:fas}:
\begin{equation}\label{eq:trop}
\phi (\tau,\zeta,\lambda) = \phi_{0} (\zeta,\lambda)  + \int_{0}^{\tau}
\Gamma_-(\tau, z;\zeta) \phi_{0} (z,\zeta,\lambda) dz
\end{equation}
We have to keep in mind also that all solutions and the scattering matrix
of the system \e{eq:L1.6} are meromorphic functions of $\lambda $.

Then we obtain that $\Gamma_-(\tau,y;\zeta) $ must satisfy the following
GLM--type equation:
\begin{equation}\label{eq:GLM}
\Gamma_-(\tau,y;\zeta)  + S_0 (\zeta) K(\tau + y;\zeta) + \int_{0}^{\tau}
 \Gamma_-(\tau,z;\zeta) K'(\tau + z;\zeta) \, dz = 0,
\end{equation}
where the kernel $K(\tau ;\zeta)  $ and its derivative $K' = dK(\tau
;\zeta) /d\tau $ are given by:
\begin{equation}\label{eq:GLMk}
K(\tau ;\zeta) = g_0 (\zeta) \left( \begin{array}{cc} 0 & k
\\ - k^* & 0  \end{array} \right) g_0^{-1}(\zeta), \qquad
k(\tau ;\zeta) = - \sum_{\lambda_j \in \cal{S}}^{} {m_j(\zeta) \over
\lambda_j} e^{i\lambda_j\tau/2}.
\end{equation}
Here $\lambda_j $ are the discrete eigenvalues of $M(\lambda) $ and
$m_j(\zeta) $ is related to the norm of the corresponding Jost solution of
\e{eq:L1.6}; generically $\lambda_j $ may depend also on $\zeta $.

The corresponding potential of \e{eq:L1.6} is recovered from the solution
$\Gamma_-(\tau,y ;\zeta)  $ of \e{eq:GLM} through:
\begin{equation}\label{eq:Ls}
S(\tau,\zeta) = B_-(\tau, \zeta) S_0(\zeta)  B_-^{-1}(\tau,\zeta) ,
\qquad B_-(\tau, \zeta)  = \bbbone + \Gamma_-(\tau, \tau ,\zeta)
S_0(\zeta) .
\end{equation}
The complete solution of the problem requires also the calculation of the
$\zeta $--dependence of the scattering data, in our case $m_j(\zeta) $ and
$\lambda_j $.

\section{Periodic, Soliton and Similarity Solutions}

  In this section we generalize the results of \cite{9,17} to describe the
similarity solutions--solitons, periodic and self-similar solutions for the
system (\ref{eq:k1q}).

We also have not been able to resolve the fundamental problem, inherent to
this type of NLEE. We have to solve the ISP for $M(\lambda) $ on a finite
interval, and naturally the $\zeta $--dependence of the corresponding
scattering data will depend on the boundary values of $Q $ at both end of
this interval. Indeed, the initial conditions  allow us to calculate all
the necessary quantities such as $S_3(\zeta,\tau =0)$, $S_\pm(\zeta,\tau=0)
$ at $\tau =0 $. In order to evaluate them at $\tau =1 $ we have to
solve completely the problem.

On the other hand, the initial conditions of such physical system must
determine uniquely its evolution. One way out of this problem is to impose
certain boundary conditions on the operator $M(\lambda) $ -- e.g.,
(quasi--) periodic. They will relate the values $S_3(\zeta,\tau =0)$,
$S_\pm(\zeta,\tau=0) $ to $S_3(\zeta,\tau =1)$, $S_\pm(\zeta,\tau=1) $.
\subsection{Periodic (cnoidal) and solitary waves}

   At first we will study the  cnoidal wave  similarity solutions
which include solitons as a special limit. If we introduce the retarded
coordinate $\xi=\zeta-\tau/\alpha$ and the following new variables
\begin{eqnarray} \label{eq:Trans1}
& &B_p(\zeta)=e^{-i\epsilon_p\tau/2\alpha} E_p(\zeta,\tau)  ,
\quad B_s(\zeta)=e^{i\epsilon_s\tau/2\alpha} E_s(\zeta,\tau)  , \nonumber \\
& &B_a(\zeta)=e^{-i\epsilon_a\tau/2\alpha} E_s(\zeta,\tau)  , \quad
Y(\zeta)=e^{-i\epsilon\zeta/2} Q(\zeta,\tau)  ,
\end{eqnarray}
where $\epsilon,\epsilon_{p,s,a}$ are arbitrary real parameters. We find that
the reduced variables $B_p$, $B_s$, $B_a$ and $Y$ satisfy:
\begin{eqnarray}\label{eq:BBY}
& &\frac{d B_p}{d\xi}=\beta_a B_{a} Y^{*}e^{-i\epsilon\xi}-
e^{i\epsilon\xi} B_s Y, \qquad
\frac{d B_s}{d\xi} = e^{-i\epsilon\xi} B_p Y^{*},  \\
& &\frac{d B_a}{d\xi} = -\beta_a e^{i\epsilon\xi} B_p Y  , \qquad
\frac{1}{\alpha}\frac{d Y}{d\xi}+e^{-i\epsilon\xi} B_p B^{*}_s
+\beta_{s} e^{-i\epsilon\xi} B_p^{*} B_a=0, \nonumber
\end{eqnarray}
where $\epsilon_p=\epsilon_s+\epsilon,  \epsilon_a=\epsilon_p+\epsilon$.
The general solution to Eq. (\ref{eq:BBY}) can be expressed in terms of
elliptic integrals. When $\epsilon=0$, the solutions may be explicitly
written in terms of cnoidal functions. Indeed, from \e{eq:k1q} and
\e{eq:k2} and using the transformed variable $\xi=\zeta-\tau/\alpha$ we
find the system:
\begin{eqnarray} \label{eq:k3}
\frac{d S_{3}}{d \xi} = -i Q^{*} S_{+}+i Q S_{-} , \quad
\frac{d S_{+}}{d \xi} = -i Q S_{3} , \quad
\alpha \frac{d Q}{d \xi} = 2i S_{+} ,
\end{eqnarray}
which has the following first integrals
%
\begin{equation}\label{eq:intk3}
\frac{1}{2} S_3^2+S_{+} S_{-} = I_1 , \qquad
S_{+} Q^{*}+S_{-} Q = I_2 , \qquad
\frac{1}{\alpha} S_{3} +\frac{1}{2} |Q|^{2} = I_3  .
\end{equation}
Introducing the new real variables $A_{+}, \phi_{+}, \tilde{Q}$ and $\phi$
by:
\begin{equation}\label{eq:AB}
S_{+}=e^{i\phi_{+}} A_{+},\qquad S_{-}=e^{-i\phi_{+}} A_{+},\qquad
Q=\tilde{Q} e^{i\phi} , \qquad \tilde{\phi}=\phi-\phi_{+} .
\end{equation}
we rewrite \e{eq:k3} and \e{eq:intk3} as follows:
\begin{eqnarray} \label{eq:k4}
& &\frac{d S_{3}}{d \xi} = -2\tilde{Q} A_{+}\sin{\tilde{\phi}}  , \qquad
\frac{d A_{+}}{d \xi} = \tilde{Q} S_{3} \sin{\tilde{\phi}} ,
\nonumber \\
& &\alpha \frac{d \tilde{Q}}{d \xi} = 2 A_{+}  \sin{\tilde{\phi}}
,\qquad \frac{d \phi}{d \xi} =  -\frac{2}{\alpha}  \frac{A_{+}}{\tilde{Q}}
\cos{\tilde{\phi}} ,  \\
\label{eq:D}
& &I_{1} = \frac{1}{2} S_{3}^2+A_{+}^2 , \qquad I_{2} = 2\tilde{Q}
 A_{+} \cos{\tilde{\phi}} , \qquad I_{3} =
\frac{1}{\alpha} S_{3}+\frac{1}{2}\tilde{Q}^2 .
\end{eqnarray}
Squaring the equation for $S_{3}$ and using the expression for $I_{k}$,
$k=1,2,3 $ we obtain
\begin{equation} \label{eq:S3}
\dot{S}_3^2 = \frac{4}{\alpha} (S_3-Z_1)(S_3-Z_2)(S_3-Z_3),
\end{equation}
where the constants $Z_i$ are related to $I_k $ by:
\begin{eqnarray}
&  Z_1 + Z_2 + Z_3 = \alpha I_3  , \qquad
 Z_1 Z_2 + Z_2 Z_3 + Z_3 Z_1 = -2 I_1 & \nonumber \\
& Z_1 Z_2 Z_3 = \alpha I_{2}^{2}/4- 2\alpha I_{1} I_{3}. &
\end{eqnarray}

   The solutions for $S_3$ may be written explicitly in terms of Jacobian
$\mbox{sn}$ function. We have the following
 periodic (cnoidal) waves:
\begin{enumerate}
\item  For $\alpha > 0$, $Z_1 < 0 < Z_2 < Z_3$,
\begin{eqnarray}
&S_3 = Z_1 + ( Z_2-Z_1 )  \mbox{sn}^2 [p(\xi-\xi_0),k)], &\\
&\displaystyle p = \sqrt{\frac{Z_3-Z_1}{\alpha}},\qquad k^2 =
\frac{Z_2-Z_1}{Z_3-Z_1}. & \nonumber
\end{eqnarray}

 \item For $\alpha < 0$, $Z_1 < Z_2 < 0 < Z_3$,
\begin{eqnarray}
&S_3 = Z_3 + ( Z_3-Z_2 )  \mbox{sn}^2 [ p (\xi-\xi_0), k) ], \\
& \displaystyle p = \sqrt{\frac{Z_3-Z_1}{-\alpha}}, \qquad  k^2 =
\frac{Z_3-Z_2}{Z_3-Z_1}.  \nonumber
\end{eqnarray}

In the particular case when $k=1 $ we find the corresponding solitary waves

\item  For $\alpha > 0$, $ Z_1 < 0 < Z_2=Z_3$,
\begin{equation}
\displaystyle
S_3 = Z_2 - \frac{Z_2-Z_1}{\cosh^2 \sqrt{ \frac{Z_2-Z_1}{\alpha} }
(\xi-\xi_0)} ,
\end{equation}

\item  For $\alpha < 0$, $Z_1=Z_2 < 0 < Z_3$,
\begin{equation}
\displaystyle
S_3 = Z_2 + \frac{Z_3-Z_2}{\cosh^2 \sqrt{ \frac{Z_3-Z_2}{-\alpha} }
(\xi-\xi_0)} .
\end{equation}
\end{enumerate}

Let us concentrate on the most important from the physical point of view
soliton solutions, i.e. the third case with additional constraint
$Z_1=Z_2=-\alpha I_3$. The result of integration is
\begin{eqnarray} \label{eq:qsol}
&\displaystyle\tilde{Q}=\frac{2 \sqrt{I_3}}{\cosh(z)} ,\qquad
S_3=\alpha I_3 (\tanh^{2}(z)-\mbox{sech}^{2}(z) ) , & \\
&\displaystyle A_{+}=\sqrt{2}\alpha I_{3}\frac{\tanh(z)}{\cosh(z)} ,
\qquad \phi_{+}-\phi=\pi/2  , \quad z=\sqrt{2 I_3} (\xi-\xi_0) .
\nonumber
\end{eqnarray}
where $\xi_{0}$ is the arbitrary initial phase. We will return again to
this solution in section 3.4.

\subsection{Self--similarity solutions}

In order to obtain the self--similarity solutions we introduce the reduced
variables
\begin{eqnarray} \label{eq:Trans2}
& &E_p(\xi)=e^{-i\epsilon_{p}\ln{\tau}} B_p(\zeta,\tau) ,\quad
E_s(\xi)=e^{-i\epsilon_{s}\ln{\tau}} B_s(\zeta,\tau) , \nonumber \\
& &E_a(\xi)=e^{-i\epsilon_{a}\ln{\tau}} B_a(\zeta,\tau) , \quad
Q(\zeta)=\frac{1}{\zeta} e^{i\epsilon\ln{\zeta}} Y(\zeta,\tau) ,
\end{eqnarray}
where $\xi=\zeta\tau$. Then the equations \e{eq:k1} becomes
\begin{eqnarray}\label{eq:20}
& &\frac{d B_p}{d\xi} = -\frac{1}{\xi} e^{i\epsilon\ln{\xi}} B_s Y
+\frac{\beta}{\xi} e^{-i\epsilon\ln{\xi}} B_a Y^{*} , \qquad
\frac{d B_s}{d\xi} = \frac{1}{\xi} e^{-i\epsilon\ln{\xi}} B_p Y^{*} ,
\nonumber \\ & &\frac{d B_a}{d\xi} = -\frac{\beta}{\xi}
e^{i\epsilon\ln{\xi}} B_p Y ,  \qquad
\frac{d Y}{d\xi}=e^{-i\epsilon\ln{\xi}} ( B_p B^{*}_s+
B_a B^{*}_p ) ,
\end{eqnarray}
where $\epsilon_{p}=\epsilon_{s}+\epsilon$,
$\epsilon_{a}=\epsilon_{p}+\epsilon$. For $\epsilon=0$ \e{eq:20}
simplifies to:
\begin{eqnarray}\label{eq:21}
& &\frac{d B_p}{d\xi} = -\frac{1}{\xi} B_s Y
+\frac{\beta}{\xi} B_a Y^{*} , \qquad
\frac{d B_s}{d\xi} = \frac{1}{\xi}  B_p Y^{*} ,   \nonumber \\
& &\frac{d B_a}{d\xi} = \frac{\beta}{\xi}
B_p Y ,  \qquad \frac{d Y}{d\xi}=B_p B^{*}_s+
\beta B_a B^{*}_p  ,
\end{eqnarray}
and, while the general solution is singular, nonsingular solutions in
terms of series can be obtained by a technique described in  \cite{9}.

We prefer here to analyze the solutions of \e{eq:k2} with another
self--similarity variable $\xi=2\sqrt{2\zeta\tau}$. If we choose
\begin{equation}\label{eq:**}
S_3=\cos[\beta(\xi)] , \qquad S_{+}=\frac{i}{2}\,\sqrt{2}\sin[\beta(\xi)]
\end{equation}
we find that \e{eq:k2} goes into:
\begin{eqnarray}\label{eq:SS}
\frac{d^2\beta (\xi)}{d\xi^2} + \frac{1}{\xi}
\frac{d\beta(\xi)}{d\xi} + \sin[\beta(\xi)] = 0,
\end{eqnarray}
Equation (\ref{eq:SS}) was first derived in another context for the
transient stimulated Raman scattering by Elgin and O' Hare \cite{19}.
This equation can be reduced to one of the standard forms of the Painleve
($P_{\rm III}$) equation \cite{20}. When $\xi \gg 1$, we can use the
asymptotic formula given by Novokshenov \cite{20} to obtain
\begin{equation}\label{eq:N}
\beta(\xi)= \frac{\tilde{\alpha}}{\xi^{1/2}}
\cos\left( \xi+\frac{\tilde{\alpha}^2}{16}\ln{\xi} + \psi\right) ,
\end{equation}
where
\begin{eqnarray}
& &\tilde{\alpha}^2=-\frac{16}{\pi}\ln[\cos(\beta_{0}/2)] , \qquad
\nonumber \\ & &\psi =\frac{2\ln{2}}{\pi}\ln[\cos(\beta_{0}/2)] +
\arg{\Gamma\left(\frac{i \tilde{\alpha}^2}{16}\right)} -\frac{\pi}{4} .
\end{eqnarray}
Here $\Gamma(x)$ is the Gamma function with a complex argument,
$\arg[\Gamma(x)]$ indicates its phase and $\beta_{0}=
\beta(\xi=0)$.  Similar expressions can be obtained for $S_3(\xi) $ and
$S_{+}$ from \e{eq:**} and for $Q$ from \e{eq:9}.

\subsection{Discussion}

   To obtain the bright solitons and to compare our results with the
ones of Kaplan {\it et al} \cite{18} we slightly generalize equations
(\ref{eq:k1}) (see for example \cite{eber,but}). The Raman quantum
transition between the lower (ground) and upper (excited) level, i.e. two
level atom is described by a $2 \times 2 $ hermitian density matrix
$\rho $ and the generalized Bloch equations \cite{eber,but}
\begin{eqnarray}\label{eq:F}
&\displaystyle \frac{\partial Q}{\partial
\tau}=\tilde{\Omega}^{*}_{R} \Delta, \qquad \frac{\partial
\Delta}{\partial \tau}=2\mbox{Re}(Q\tilde{\Omega}_{R}),
\qquad   \Delta = \rho_{11} - \rho_{22} , \nonumber \\
&\displaystyle \tilde{\Omega}_{R}=\frac{2}{\hbar} \left(\alpha_{s,p}
E_{s} E^{*}_{p}+\alpha_{p,a} E_{p} E^{*}_{a}\right), \qquad
Q = - 2i \rho_{12} e^{-ik_0 z + i\omega_0 t }. &
\end{eqnarray}
Here $\tilde{\Omega}_{R}$ is the generalized local Rabi frequency
\cite{eber}, $k_{0}=k_{p}-k_{s}\sim \omega_{0}/c$  and we have assumed
that $\rho_{11}+\rho_{22}=1$.

The generalization of \e{eq:k1} we mentioned above is obtained by
replacing the equation for $Q $ in \e{eq:k1} by the system \e{eq:F}.
These equations, rewritten for the self--similarity variable $\xi = \zeta
- \tau /\alpha $ coincide with equation (6) from Kaplan {\em et al.}
\cite{but}. The direct comparison of $S_{+}$ and $\tilde{\Omega}_{R}$
shows that the physical interpretation of $S_{+}$ is the normalized local
Rabi frequency.  From these equations we obtain also that the quantity
\begin{eqnarray}
\Delta^2 + |Q|^2=C(\zeta),
\end{eqnarray}
is conserved in $\tau$ and that the following equations
\begin{eqnarray}\label{eq:25a}
&\displaystyle \delta_{p} \Phi_{p}+\delta_{s} \Phi_{s}
+\delta_{a} \Phi_{a}=I(\tau),  \qquad
\delta_{i}=\left(\frac{1}{v_{gi}}-\frac{1}{\tilde{v}_{g}}\right),\quad
i=p,s,a, & \\
\label{eq:25}
&\displaystyle  \frac{\partial}{\partial \zeta}
(\delta_{a} \Phi_{a}-\delta_{s} \Phi_{s})-
\pi N_{0}\frac{\partial}{\partial \tau} \Delta = 0 ,\quad \Phi=|E_{i}|^2
&
\end{eqnarray}
hold. By $N_{0}$ we have denoted the density of Raman particles. Let us
use the following ansatz \cite{18}
\begin{eqnarray}\label{eq:M}
& &\Phi_{p}=|a_{p}|^2 \Phi_{\Sigma}, \quad
\Phi_{s}=|a_{s}|^2 \Phi_{\Sigma}, \quad
\Phi_{a}=|a_{a}|^2 \Phi_{\Sigma}, \nonumber  \\
& &|a_{p}|^2=\frac{\gamma_{3}^{2}}{W}, \quad
|a_{s}|^2=\frac{|w_{s,p}|^2}{\delta_{s}^2 W}, \quad
|a_{a}|^2=\frac{|w_{p,a}|^2}{\delta_{a}^2 W}, \\
& &w_{s,p}=\frac{2\pi}{c}\alpha_{s,p} \sqrt{\frac{\omega_{s} \omega_{p}}
{n_{s} n_{p}} }, \quad
w_{p,a}=\frac{2\pi}{c}\alpha_{p,a} \sqrt{\frac{\omega_{p} \omega_{a}}
{n_{p} n_{a}} }, \nonumber
\end{eqnarray}
where \cite{but}
\begin{eqnarray}\label{eq:alfa-sp}
\alpha_{s,p} &=&\frac{1}{\hbar^2} \sum_{m}\left[ {
(\vec{d}_{1m}.\vec{e}_{p})
(\vec{d}_{m2}.\vec{e}_{s}) \over (\omega_{m1}-\omega_{p})} +
{ (\vec{d}_{1m}.\vec{e}_{s})
(\vec{d}_{m2}.\vec{e}_{p}) \over (\omega_{m1}+\omega_{s})} \right], \\
\label{eq:alfa-pa}
\alpha_{p,a} &=&\frac{1}{\hbar^2} \sum_{m}\left[
{ (\vec{d}_{1m}.\vec{e}_{a})
(\vec{d}_{m2}.\vec{e}_{p}) \over (\omega_{m1}-\omega_{a})} +
{ (\vec{d}_{1m}.\vec{e}_{p})(\vec{d}_{m2}.\vec{e}_{a}) \over
(\omega_{m1}+\omega_{p})} \right] .
\end{eqnarray}
Here $\vec{d}_{1m},\vec{d}_{2m}$ are the dipole matrix elements between
the Raman quantum levels and the $m$-th quantum level, $m\neq 1,2$ and
$n_{p}, n_{s}, n_{a}$ are refractive indexces at frequencies $\omega_{p},
\omega_{s}, \omega_{a}$ respectively.
{}From \e{eq:alfa-sp}, \e{eq:alfa-pa} we find that if $\omega_p \gg
\omega_a, \omega_s $ and $\omega_a, \omega_p \ll \omega_{m1} $ then
$\alpha_{s,p} \simeq \alpha_{p,a} $ and $\omega_{s,p} \simeq \omega_{p,a}
$. Using arguments analogous to the ones in~\cite{1} we find that one can
expect such physical systems to be described by the system \e{eq:k1} with
$\beta_a \simeq 1 $.

Inserting \e{eq:M} into $I $
\e{eq:25a}, from \e{eq:25} we obtain:
\begin{eqnarray}
\gamma_{3}^{2}=-\left(\frac{|w_{s,p}|^{2}}{\delta_{s} \delta_{p}}+
\frac{|w_{p,a}|^{2}}{\delta_{a} \delta_{p}}\right) . \nonumber
\end{eqnarray}

Let us consider the second integral \e{eq:25} with conserved density
$\delta_{a} \Phi_{a}-\delta_{s} \Phi_{s}$ and  conserved flux --
$-\pi N_0\Delta$. Then
\begin{equation}\label{eq:J}
J = 2 \left( \delta_a \Phi_a  - \delta_s \Phi_s  \right)
= -{ 2\pi \Phi_\Sigma N_0 \over \Phi_0}.
\end{equation}
Here
\begin{eqnarray}\label{eq:F0}
\Phi_{0}=\frac{\pm\pi N_{0} W}
{ ( |w_{s,p}|^2/\delta_{s}-|w_{p,a}|^2/\delta_{a}) } ,
\end{eqnarray}
where "$-$" indicates that the molecules (atoms) are initially at the
equilibrium and "$+$" that the population difference is inversed. Let us
also introduce:
\begin{eqnarray}\label{eq:F1}
\Phi_{\Sigma}=\Phi_{0} S(\xi),\qquad \Delta=\pm (1-S(\xi)),\qquad
Q(\xi)=-\gamma_{3}\xi S(\xi),
\end{eqnarray}
where $S(\xi)$ may have different forms (soliton, Lorentzian etc.) \cite{18}.
Finally from the normalization condition $|a_{p}|^2+|a_{s}|^2+|a_{a}|^2=1$
we have
\begin{eqnarray}
W=\frac{|w_{s,p}|^2}{\delta_{s} \delta_{s,p}}-
  \frac{|w_{p,a}|^2}{\delta_{a} \delta_{p,a}}  ,
\end{eqnarray}
where
\begin{eqnarray}
\delta_{s,p}= \frac{1}{\delta_{s}}-\frac{1}{\delta_{p}} , \qquad
\delta_{p,a}= \frac{1}{\delta_{p}}-\frac{1}{\delta_{a}} .
\end{eqnarray}
Recently these bright solitons, in more general physical situation,
cascade SRS \cite{1,but} have been used to predict generation of
subfemtosecond coherent pulses in SRS experiment \cite{18}. From the
above analysis it is clear, that bright solitons are obtained in the
case of finite group velocity dispersion parameters $\delta_{i}$
(\ref{eq:25}) \cite{1,18} and non--zero population difference $\Delta$.

\subsection{The one soliton solution}
Here we will show, that the auxiliary linear problem for the vector NLS
equation with some additional reduction is equivalent to the
Stokes--anti--Stokes SRS equations without last equation for $Q$.
This formal equivalence allows us to recover $E_{p}, E_{s}, E_{a}$ from
potential $Q$, already obtained by inverse scattering transform method of
Sec.~2 with \e{eq:9}.

Indeed, if we introduce:
\begin{eqnarray}\label{eq:4.1}
{\partial  \over \partial  \zeta} \left( \begin{array}{c} \tilde{\psi}_1
\\ \tilde{\psi}_2 \\ \tilde{\psi}_3 \end{array} \right) =
\left( \begin{array}{ccc}  0  & q_1 & q_2 \\
-q_1^*, & 0 & 0 \\ - q_2^* & 0 & 0
 \end{array} \right) \left( \begin{array}{c} \tilde{\psi}_1
\\ \tilde{\psi}_2 \\ \tilde{\psi}_3 \end{array} \right)
\end{eqnarray}
and require that:
\begin{equation}\label{eq:4.2}
q_1 = -Q, \quad q_2 = Q^*, \qquad \psi_{1} = E_p, \quad
\psi_{2} = E_s,
\quad \psi_{3} = E_a.
\end{equation}
without last equation for $Q$, which we may obtain from the previous
subsection. Using the well known one soliton solution of the vector nonlinear
Schr\"{o}dinger equation under the reduction (\ref{eq:4.2}) and the
solution of the linear problem (\ref{eq:4.1}) with potentials $Q, Q^{*}$ we
obtain
\begin{eqnarray} \label{eq:sol1}
Q=\frac{\sqrt{2}\eta e^{i\phi}}{\cosh(z)}  ,\quad
z=\eta \zeta-\frac{1}{\eta}\int_{0}^{\tau} K^{2}(\tau') d\tau'
\end{eqnarray}
where $K(\tau')$ is real, the soliton's eigenvalue is $i\eta$ and $\phi$ is
constant real phase. The direct integration
of Eq. (\ref{eq:4.1}) with reduction (\ref{eq:4.2}) is given by
\begin{eqnarray} \label{eq:sol2}
E_{p}=\sqrt{2} K(\tau) \frac{\tanh(z)}{\cosh(z)} e^{i\phi} ,\,\,
E_{s}=K(\tau) \tanh^{2}(z), \,\,
E_{a}=\frac{K(\tau)}{\cosh^{2}(z)} e^{2i\phi} .
\end{eqnarray}
These solutions are similar to transient SRS solitons obtained in \cite{9}
(see also \cite{21}) and for $K=1, \phi=0$ coincide with the ones in
\cite{1}. If we now calculate the $S_{3}$ and $S_{\pm}$ using the above
expressions for $E_p, E_s, E_a$ we obtain precisely the soliton solution
(\ref{eq:qsol}).

\section{ $N$--component generalizations}

   In this section we show that the extended model with $N$--Stokes and
$N$--Stokes components is also integrable in the sense means of IST method.
The considerations are formal from physical point of view.

Let us consider the following equations
\begin{equation}\label{eq:5.1}
\begin{array}{lll} \displaystyle
\frac{\partial E_p}{\partial \zeta}=
\sum_{i=1}^{N}(\beta_a Q^{*} E_a^{(i)}-Q E_s^{(i)}),  &  &
\displaystyle \frac{\partial E_s^{(i)}}{\partial \zeta} = Q^{*} E_p,
 \qquad  i=1,\ldots, N  , \\[5pt]
\displaystyle \frac{\partial E_a^{(i)}}{\partial \zeta}  = -\beta_a Q E_p,
&   & \displaystyle
\frac{\partial Q}{\partial \tau} + \tilde{g} Q=
\sum_{i=1}^{N}(E_s^{(i) *} E_p+\beta_a  E_p^{*} E_a^{(i)}).
\end{array}
\end{equation}
with $\beta_a = 1$.
 The equations for the $E_{k}^{i}$ can be written down as the same
auxiliary linear problem, which  solves the $N$--component vector
NLS equation
\begin{eqnarray}
& & q_1 = -Q, \qquad   q_2 = Q^*, \quad  \dots \quad
    q_{2N-1} = -Q, \qquad   q_{2N} = Q^*, \nonumber \\
& &\psi_{1} = E_p, \quad
\psi_{2} = E_s^{(1)}, \nonumber \\
& &\psi_{3} = E_a^{(1)}, \quad
  \dots  \quad \psi_{2N} = E_s^{(N)}, \qquad
\psi_{2N+1} = E_a^{(N)}, \nonumber
\end{eqnarray}
\begin{eqnarray}
{\partial  \over \partial  \zeta} \left( \begin{array}{c} \tilde{\psi}_1
\\ \tilde{\psi}_2 \\ \tilde{\psi}_3 \\
\vdots \\ \tilde{\psi}_{2N} \\ \tilde{\psi}_{2N+1}
 \end{array} \right) = \left( \begin{array}{cccccc}  0 &
q_1 & q_2 & \dots & q_{2N-1} & q_{2N}\\ -q_1^* & 0 & 0 &
 \dots & 0 & 0 \\ - q_2^* & 0 & 0 &\vdots & 0 & 0 \\
 \vdots & \vdots & \vdots & \ddots & \vdots & \vdots \\
 - q_{2N-1}^* & 0 & 0 & \dots & 0 & 0 \\
 - q_{2N}^* & 0 & 0 & \dots & 0 & 0
\end{array} \right)
 \left( \begin{array}{c} \tilde{\psi}_1
\\ \tilde{\psi}_2 \\ \tilde{\psi}_3 \\
\vdots \\ \tilde{\psi}_{2N} \\ \tilde{\psi}_{2N+1} .
 \end{array} \right) \nonumber
 \end{eqnarray}
with the spectral parameter $\lambda =0 $.

We introduce again the quadratic variables
\begin{eqnarray}\label{eq:5.2}
  S_3 = \frac{1}{2} \sum_{i=1}^{N} ( |E_s|^2 - |E_a|^2 ), \qquad
S_{+} = S_{-}^{*} = \frac{i}{2} \sum_{i=1}^{N} ( E_s^{*} E_p
+ E_p^{*} E_a ),
\end{eqnarray}
and show that they satisfy the same equation \e{eq:k2}.
Therefore the ZS--AKNS representation (\ref{eq:ZS}) can be used as above
for the analysis of the system (\ref{eq:5.1}). The procedure of solving
Eq. (\ref{eq:5.1}) is analogous to the considerations of section IV.
Clear physical interpretation and solutions of Eq. (\ref{eq:5.1}) will be
given elsewhere.

\section{Conclusion}

   In this paper a method of solving Stokes-anti Stokes SRS with $\beta_a
= 1$ is presented.  New transient solitons (\ref{eq:sol1}),
(\ref{eq:sol2}) are obtained. For bright solitons our results are in
agreement with these of Kaplan {\it et al} \cite{18}. The traveling and
self-similar solutions to Eq.~(\ref{eq:k1}) are discussed.

We stress also that for the transient solitons with $\beta_a=1 $ the
number of Stokes--anti--Stokes photons are close to each other.

The ISM generically meets with difficulties due to the fact that
generically no boundary conditions on the potentials \e{eq:L1.6} are
imposed. In the case of quasiperiodic boundary conditions these
difficulties are overcome.

Looking at the Lax operator \e{eq:L1.6} we recognize the system of
equations \e{eq:k2} as belonging to the Heisenberg ferromagnet hierarchy.
One can try to apply the expansions over the ``squared solutions'' in the
spirit of~\cite{VSG} and then treat the case $\tilde{g} \neq 0 $ as a
perturbation.

\section{ Acknowledgements }

    We are grateful to Prof. D. J. Kaup and to Drs. L. Kovachev and I.
Uzunov for helpful discussions.  One of the authors (VSG) has been
partially supported by contract F-215 with the Ministry of Science the
other (NAK) acknowledges support from EEC grant ERB-CIPA-CT-92-0473.

\end{document}